\documentclass[12pt]{article}

\usepackage[margin=1in]{geometry} 
\usepackage{amsmath,amsthm,amssymb}
\usepackage{textcomp}
\usepackage[dvips,final, pdftex]{graphicx}         		
\usepackage[greek, english]{babel}                			
\usepackage[hang,bf]{caption}              			
\usepackage{rotating}                     		 		
\usepackage{subfigure}
\usepackage{units}
\usepackage{fancyhdr}
\usepackage{lscape}
\usepackage{amsfonts}
\usepackage{float}
\usepackage{graphicx}
\usepackage{longtable}
\usepackage{fancybox}
\usepackage[hidelinks]{hyperref}
\usepackage[capitalise,nameinlink]{cleveref}
\usepackage{here}
\usepackage{setspace} 
\usepackage{units} 
\usepackage{pdfpages}
\usepackage{lscape} 		
\usepackage{pdflscape}
\usepackage{epstopdf}
\usepackage{epsfig}
\usepackage{soul}

\usepackage{upgreek}

\usepackage{listings}

\newcommand{\dd}{\mathrm{d}}

\begin{document}
	\bibliographystyle{elsarticle-num}
	\title{Explicit and asymptotic solutions for frictional incomplete half-plane contacts subject to general oscillatory loading in the steady-state}

	\author{H. Andresen$^{\,\text{a,}}$\footnote{Corresponding author: \textit{Tel}.: +44 1865 273811; \newline \indent \indent \textit{E-mail address}: hendrik.andresen@eng.ox.ac.uk (H. Andresen).}, R.M.N. Fleury$^{\,\text{b}}$, M.R. Moore$^{\,\text{c}}$, D.A. Hills$^{\,\text{a}}$ $\,\,$\\ \\
		\scriptsize{$^{\text{a}}$ Department of Engineering Science, University of Oxford, Parks Road, OX1 3PJ Oxford, United Kingdom}\\
		\scriptsize{$^{\text{b}}$ Institut f\"ur Baumechanik und Numerische Mechanik, Leibniz Universit\"at Hannover, Appelstr. 9A, 30167 Hanover, Germany}\\
		\scriptsize{$^{\text{c}}$ Mathematical Institute, University of Oxford, Andrew Wiles Building, Radcliffe Observatory Quarter,}\\
		\scriptsize{Woodstock Road, OX2 6GG Oxford, UK}
	}
%	
%	\author{H. Andresen$^{\,\text{a,}}$\footnote{Corresponding author: \textit{Tel}.: +44 1865 273811; \newline \indent \indent \textit{E-mail address}: hendrik.andresen@eng.ox.ac.uk (H. Andresen).}, R.M.N. Fleury$^{\,\text{b}}$, M.R. Moore$^{\,\text{c}}$, D.A. Hills$^{\,\text{a}}$ $\,\,$\\ \\
%				\scriptsize{$^{\text{a}}$ \textit{Department of Engineering Science, University of Oxford, Parks Road, OX1 3PJ Oxford, United Kingdom}} \\
%				$\,$\scriptsize{$^{\text{b}}$ \textit{Institut f\"ur Baumechanik und Numerische Mechanik, Leibniz Universit\"at Hanover,}}\\
%		\scriptsize{\textit{Appelstra\ss e 9a, 63167 Hannover, Germany}} \\
%			$\,$\scriptsize{$^{\text{c}}$ \textit{Mathematical Institute, University of Oxford, Andrew Wiles Building, Radcliffe Observatory Quarter, }}\\
%		\scriptsize{\textit{Woodstock Road, OX2 6GG Oxford, UK}}}

	\date{}
	\maketitle
	\begin{center}
		\line(1,0){470}
	\end{center}
	\begin{abstract}
		\footnotesize{	
			This contribution presents an asymptotic formulation for the stick-slip behaviour of incomplete contacts under oscillatory variation of normal load, moment, shear load and differential bulk tension. The asymptotic description allows us not only to approximate the size of the slip zones during the steady-state of a cyclic problem without knowledge of the geometry or contact law, but provides a solution when all known analytical solutions for incomplete contacts reach their limitations, that is, in the presence of a varying moment and a differential bulk tension large enough to reverse the direction of slip at one end of the contact. An insightful comparison between the mathematically explicit analytical solution and the asymptotic approach is drawn using the example geometry of a shallow wedge. \\
			}

	\end{abstract}
	\begin{center}
		\line(1,0){470}
	\end{center}

\section{Introduction}\label{Introduction}

\begin{figure}[t!]
	\centering
	\includegraphics[scale=0.7, trim= 0 0 0 0, clip]{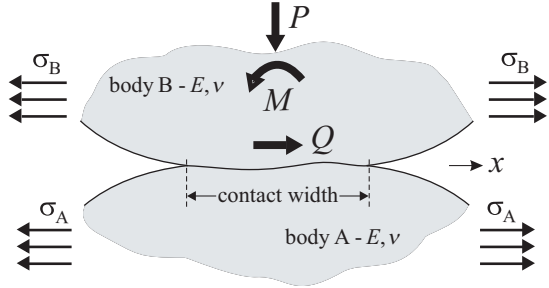}
	\caption{General half-plane contact subject to a normal load, $P$ and a  moment, $M$, with a differential tension, $\sigma = \sigma_{\text{A}} -  \sigma_{\text{B}}$.}
	\label{fig:GeneralContactPorblem}	
\end{figure}

\hspace{0.4cm}The problem we wish to address is the partial slip behaviour of a
contact, which may be of general form, and subject to a normal load, $P$,
moment, $M$, shear force, $Q$, and differential bulk tension, $\sigma$, see Figure \ref{fig:GeneralContactPorblem}. These quantities vary with time so that each has an oscillatory component and at least the normal load remains positive throughout the cyclic loading. Our practical motivation
is to be able to understand how a number of real-world
%practical
contacts respond. Examples include the 
%dovetail 
flanks of fan blade dovetail roots in gas turbines and the locking segments employed in securing the riser connector in position on a seabed wellhead.

Problems of this kind fall into two categories.%kind; 
When the differential bulk tension, $\sigma$,
is moderate, meaning the slip zones at each side of the contact are of the same
sign, a method based on superposition of the full
sliding traction, and whose lineage goes right back to the original
paper \cite{Cattaneo_1938} may be used. Cattaneo showed that, for
a Hertzian contact under constant normal load the sliding shear traction
produced a linearly varying surface strain so that, by superposing
a second, scaled form of the shear traction of opposite sign the relative slip displacement in the central, stick region is made zero,
as required. J\"ager and Ciavarella later demonstrated that the same idea
might be applied to a contact of any geometry \cite{Jaeger_1997}, \cite{Ciavarella_1998}. Nowell and Hills found that, for a Hertzian contact under constant normal load, in the presence of moderate tension the stick zone would be
offset, and a similar superposition might again be used \cite{Hills_1987}. A major
%big 
breakthrough came when Barber et al. \cite{Barber_2011} showed that a contact which traced out an a convex loop in normal force-shear force space might also be treated by a variation on the same ideas.

\begin{figure}[t!]
	\centering
	\includegraphics[scale=0.7, trim= 0 0 0 0, clip]{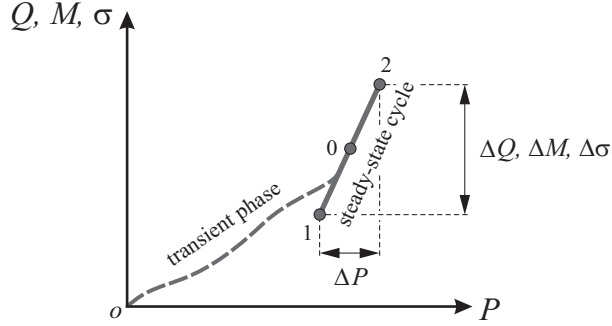}
	\caption{Cyclic loading with in phase changes in all load components.}
	\label{fig:CyclicLoop}	
\end{figure}

\begin{figure}[b!]
	\centering
	\includegraphics[scale=0.33, trim= 0 0 0 0, clip]{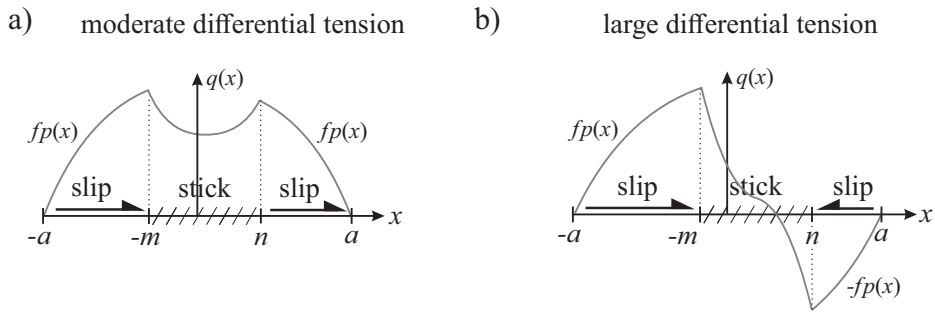}
	\caption{Illustration of a steady-state cycle with a proportionally varying normal load, moment, shear load and differential bulk tension.}
	\label{fig:ModerateLargeTension}	
\end{figure}

Here, we do not wish to consider quite such a general problem in terms
of the possible phase shift between the two components of load because
we are interested in practical problems where one applied load (the
centrifugal force in a gas turbine, the hydraulic locking pressure
in a wellhead connector) excites all four components of local contact
load but is then held constant while a second load (vibration in the
case of a gas turbine, surface waves moving the riser in the case
of the connector) induce changes again in all four elements of contact
load but, crucially, \emph{in phase}. Figure \ref{fig:CyclicLoop} shows a generic steady-state cycle between the maximum (2) and minimum (1) points, 
%$2$ -- $1$ 
under proportional loading. The $\Delta$-terms indicate the range of the individual load components with the mean point at $0$. Before proceeding to the partial slip problem, we start by writing down the conditions for a contact to remain fully adhered due to small changes in the load quantities \cite{Andresen_2019_3} 
\begin{equation}\label{fullstickinequality}
\frac{4a \,\dd Q\pm\pi a^{2}\,\dd\sigma}{4a\,\dd P\pm 8\,\dd M}<f \; \text{,}
\end{equation}
where $a$ is the instantaneous contact half-width, $f$ is the coefficient of friction, and where we choose the upper signs when considering the left-hand side of the contact and the lower signs when considering the right-hand side of the contact. We expect microscopic amounts of differential relative motion to ensue when the inequality in Eq. \eqref{fullstickinequality} becomes violated. Figure \ref{fig:ModerateLargeTension} a) illustrates shear tractions in partial slip with regions of slip and the extent of the steady-state stick zone is denoted by $[-m\qquad n]$. We consider that, once the steady-state permanent stick zone in a partial slip problem is known other aspects of the solution, such as the relative surface slip displacement, will fall readily into place. Note that the transient phase has no effect on the extent and position of the steady-state permanent stick zone, 
%spanning $[-m \qquad n]$ 
but merely on the locked-in tractions, and hence the transient is irrelevant to the solution we seek. In a recent series of articles we have shown how superposition of the sliding tractions, $f p(x)$ may be applied to contact problems which are symmetrical but subject to a change in a normal load (but no moment) \cite{Andresen_2019_2}, and unsymmetrical contacts allowing for the possibility of a moment \cite{Andresen_2019_3}. We emphasize that this whole family of solutions will apply only when the sign of slip is the same at each edge of the contact, i.e. the differential tension is moderate.

When the bulk tension developed is large, see Figure \ref{fig:ModerateLargeTension} b), it reverses the sign of slip
at one contact edge and therefore all of the methods described above
do not apply, so a different approach is needed. One emerged when
a solution was found for the state of shear traction developed with a glide
dislocation present on the interface between two half-planes joined
over an interval equal to the size of the contact \cite{Moore_2018}. This made it possible to start with the opposite assumption - that the contact was fully stuck - and then to use the solution for the dislocation as a kernel to add the
regions of slip. This approach works well but it is necessary to know
the shear traction change under fully stuck conditions, as the contact
moves from one turning point in the loading cycle to the other \cite{Andresen_2019_4}. The problem comes when a varying moment
is part of the load cycle as this means that it is not possible to
obtain a simple recipe to give the shear traction change needed (under
conditions of full-stick) \cite{Andresen_2020}.

These difficulties mean that attention must be turned to each edge
of the contact, in turn, and a solution sought within the framework
of an asymptotic formulation, together with the contact law. This
paper describes, in detail, how this may be done and makes a comparison with the true answer \cite{Andresen_2019_4} and \cite{Moore_2020}
for the case of a shallow wedge contact geometry where applicable.

\section{Mossakovskii Methods} \label{mossakovskii_methods}

\hspace{0.4cm}One of the key components of the asymptotic methods we describe in detail in \textsection \ref{asymptotic_methods} concerns the behaviour of the contact pressure at the edges of an incomplete contact. As is well known, the contact pressure is square-root bounded at the edges of an incomplete contact \cite{Barber_2010}, and it is the coefficient, $K_{N}$, of this square-root term that we require in our asymptotic approach. Even when the contacting bodies are elastically similar so that the normal and tangential contact problems decouple, the form of $K_{N}$ is geometry dependent, and there are several approaches we could take to isolate it. However, recent work \cite{Moore_2020} using the Mossakovskii method \cite{Mossakovskii_1953} lends itself particularly well to finding $K_{N}$ for general geometries including the effects of both an applied normal force and an applied moment.

\begin{figure}[b!]
	\centering 
	\includegraphics[scale=1.0, trim= 0 0 0 0, clip]{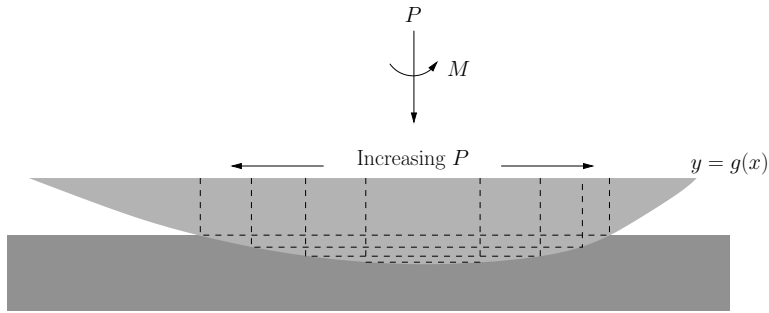}
	\caption{A large, almost flat body $y = g(x)$ is pressed into an elastically-similar half-space with an applied normal force, $P$, and an applied moment, $M$. The Mossakovskii idea approximates the geometry of the indenter by a series of flat punches, whose widths increase as the contact pressure increases.}
	\label{fig_MossakovskiiMethod} 
\end{figure}

We consider the contact of an indenter with body profile given by $y = g(x)$, where $(x,y)$ are Euclidean coordinates centred at the minimum of the body. Under an applied normal force, $P$, and an applied moment, $M$, the indenter presses into an elastically-similar half space, with the contact extending across $[-a \qquad c]$, where $a,c > 0$. The standard manner of treating this contact is then to invert a singular integral equation relating the contact pressure $p(x)$ to the body geometry gradient, see \cite{Barber_2010} for details. In doing so, one finds that a consistency condition must be satisfied for the contact pressure to be bounded
\begin{equation}
0 = \int_{-a}^{c}\frac{g'(s)}{\sqrt{(a-s)(s+c)}}\,\mbox{d}s. \label{eqn:consistencyab}
\end{equation}
The main drawback of this approach is the difficulty in dealing with the Cauchy principal value integrals that arise, since they have singular kernels within the contact. The Mossakovskii method aims to circumvent this problem by converting the singular integral form to a regular --- albeit non-symmetric --- Abel integral equation.

The central idea of the Mossakovskii method is depicted in Figure \ref{fig_MossakovskiiMethod}. The aim is to approximate the geometry of the indenter by an infinite series of flat punches. In order to do this, \cite{Moore_2020} formulate the problem in such a way that (\ref{eqn:consistencyab}) is used as an equation to isolate the left-hand contact point as a function of the right-hand contact point, i.e. giving $a(c)$. The contact pressure, $p(x,c)$, is then given through a superposition of these of the form
\begin{equation}
p(x,c) = \int_{0}^{c} F(s)h(x,s)\,\mbox{d}s,
\label{eqn:PressureSuperposition}
\end{equation}
where $h(x,a)$ is the contact pressure corresponding to the \emph{complete} contact of a flat punch, that is
\begin{equation}
h(x,c) = \begin{cases}
\displaystyle{\frac{1}{\sqrt{(a(c)+x)(c-x)}}} & \mbox{for} \; -a(c)<x<c \\
\displaystyle{0} & \mbox{otherwise}
\end{cases}.
\label{eqn:FlatPunchCP}
\end{equation}

The unknown function $F(\cdot)$ in (\ref{eqn:PressureSuperposition}) can then be shown to satisfy the non-symmetric Abel equation
\begin{equation}
\frac{E^* g'(x)\pi}{2} = \int_{0}^{x} \frac{F(s)}{\sqrt{(x-s)(x+a(s))}}\,\mbox{d}s \; \mbox{for} \; 0<x<c \; \text{,} \label{eqn:FIE}
\end{equation}
where $E^* = E/(2 (1-\nu^2))$ is the composite Young's modulus of the contacting bodies\footnote{Note that we assume the contacting bodies to be elastically similar, i.e. $(E, \nu) = (E_A, \nu_A) = (E_B, \nu_B)$, in order to uncouple the normal and tangential problem \cite{Dundurs_1969}}. One key point noted in \cite{Moore_2020} is that in order for the Mossakovskii approach to work, the minimum of the indenter must be fixed as $P$ increases, which precludes a change in $M$ as $P$ varies. To get around this, \cite{Moore_2020} assume that the geometry is known \emph{a priori} and then find the required $M$ necessary to sustain the contact for a given $P$. This is equivalent to choosing a particular path in the $(P,M)$ load space. But, as is well-known, the contact pressure --- and hence, the $K_{N}$ coefficients --- at load state $(P,M)$ is history-independent, so that this requirement is not overly-restrictive.

To find the $K_{N}$ coefficients, it transpires that we do not even need to solve (\ref{eqn:FIE}) for $F(\cdot)$. Vertical equilibrium demands that
\begin{equation}
P = \int_{-a(c)}^{c} p(s,c)\,\mbox{d}s,\label{eqn:NormalEquilibrium}
\end{equation}
which can be manipulated to show that
\begin{equation}
F(c) = \frac{1}{\pi}\frac{\mbox{d}P}{\mbox{d}c}. \label{eqn:PFRelation}
\end{equation}

Then, \cite{Moore_2020} expand (\ref{eqn:PressureSuperposition}) close to the edges of the contact and show that

\begin{equation}
p \sim \frac{2}{\pi}\frac{P'(c)}{\sqrt{c+a(c)}}\sqrt{c-x} \; \mbox{as} \; c-x\rightarrow 0, \; \mbox{so that} \; K_{N,c} = \frac{2}{\pi}\frac{P'(c)}{\sqrt{c+a(c)}}, \label{eqn:Knc}
\end{equation}
while
\begin{equation}
p \sim \frac{2}{\pi}\frac{P'(c)}{a'(c)\sqrt{c+a(c)}}\sqrt{a(c)+x} \; \mbox{as} \; a(c)+x\rightarrow 0, \; \mbox{so that} \; K_{N,a} = \frac{2}{\pi}\frac{P'(c)}{a'(c)\sqrt{c+a(c)}}.\label{eqn:Kna}
\end{equation}

\section{Asymptotic Methods} \label{asymptotic_methods}

As mentioned above, the analytical solution of the contact tractions and slip zone size can be obtained for some specific contact geometries under varying normal, moment, shear and differential bulk tension loading. However, for problems subject to large tension and a varying moment, as explained in \textsection 1, the use of an asymptotic approach, which focuses on one contact edge and its surroundings only, is adopted. Even for cases where an analytical solution exists, it might be of interest to have a succinct and easy-to-apply method at hand and this is what we shall establish here.

The asymptotic approach to incomplete contacts was first introduced in \cite{sackfield2003application} and later extended to the tangential contact problem in \cite{dini2004bounded, dini2005comprehensive}, using the Ciavarella-J\"ager theorem \cite{Ciavarella_1998,Jaeger_1998}. Asymptotic approaches have since been used for different contact problems, such as flat and rounded punches \cite{fleury16a,fleury16b}, and incomplete problems with constant normal but varying shear and bulk loading. More recently, the asymptotic approach was extended to the varying normal load case \cite{fleury2017varying}, as an asymptotic version of the work in \cite{Barber_2011}.

\begin{figure}[tbp]
	\centering
	\includegraphics[width=0.6\linewidth]{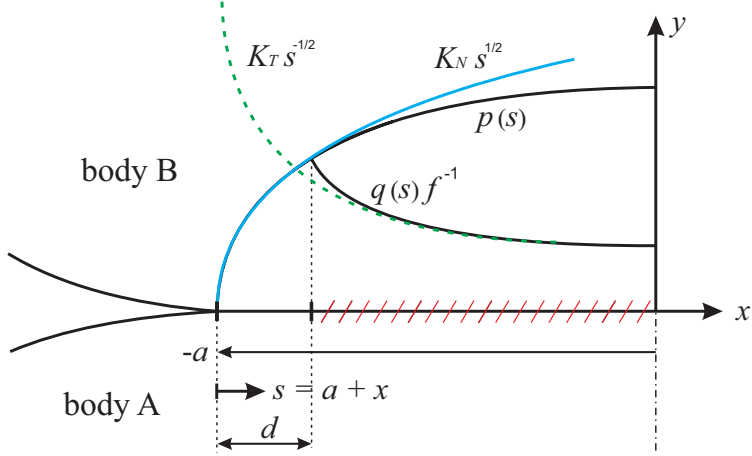}
	\caption{Normal and tangential asymptotes at the edge of an incomplete contact.}
	\label{fig:asymptotes}
\end{figure}

In the tangential contact problem, the shear tractions are split between the sliding solution, which is similar to the contact pressure but scaled by the coefficient of friction, and the stick solution. The former is bounded for incomplete contacts, as is the contact pressure, and the latter is singular when all slip is inhibited throughout the contact interface. In the asymptotic approach, we focus our interest on the edges of the contact, where here we will present results for the left-hand edge $x = -a$ without loss of generality. We shall retain the global axis set centred at the body minimum to permit the effect of the moving contact edge to be incorporated readily in the analysis. As discussed in \textsection \ref{mossakovskii_methods}, local to $x = -a$, the pressure may be expanded as an asymptotic series. In this analysis, we shall simply take the first term in this expansion and discard the higher-order terms. We can similarly expand the shear traction, $q(x)$, and we shall again only retain the first, inverse square-root term in the expansion. Thus, local to $x = -a$, we have the asymptotic approximations
\begin{equation}
p(x) \approx K_N \sqrt{a+x}, \; q(x) \approx \frac{K_{T}}{\sqrt{a+x}}.
\label{eq:normalasymptote}
\end{equation}
These asymptotes are illustrated in Figure \ref{fig:asymptotes} alongside sketches of the exact forms of $p(x)$ and $q(x)$.

The normal and tangential asymptotes are represented by the coefficients $(K_{N},K_{T})$, which are analogous to generalised stress intensity factors. The coefficient $K_{N}$ for the asymptote approximating the normal contact pressure at the edges of the contact has been explicitly obtained in (\ref{eqn:Knc})--(\ref{eqn:Kna}) for each edge of the contact. Note that the $K_N$ coefficients are geometry dependent.
The coefficients of the tangential contact asymptotes, $K_{T}$, on the other hand, are geometry independent and may be obtained from the shear load, $Q$, and differential bulk tension, $\sigma$, as in \cite{hills16}
\begin{equation}
K_T = \frac{\pm Q}{\pi \sqrt{2 a}} + \frac{\sigma}{4} \sqrt{\frac{a}{2}}.
\end{equation}%

%%%%%%%% %%%%%%%% %%%%%%%% %%%%%%%% %%%%%%%% %%%%%%%% %%%%%%%% %%%%%%%% %%%%%%%% 
%%%%%%%% %%%%%%%% %%%%%%%% %%%%%%%% %%%%%%%% %%%%%%%% %%%%%%%% %%%%%%%% %%%%%%%% 
%%%%%%%% %%%%%%%% %%%%%%%% %%%%%%%% %%%%%%%% %%%%%%%% %%%%%%%% %%%%%%%% %%%%%%%% 

\subsection{Asymptotic approximation of shear tractions}

The shear traction at the maximum shear load during a generic steady-state loading cycle, as depicted by point $2$  in Figure \ref{fig:CyclicLoop}, can be expressed as a scaled superposition of the contact pressure at different load pressures throughout the loading path, i.e.~using the  Ciavarella-J\"ager theorem, as in \cite{Barber_2011}
\begin{equation} 
q_1(x) = \int_{0}^{P_Z} \frac{\partial p(x,P,M)}{\partial P} \frac{\dd P}{\dd Q} \dd P + fp(x,P,M) - 2 fp(x, P_S,M_S) + fp(x,P_Z,M_Z) \; \text{,}
\label{eq:CJshear}
\end{equation}%  
where $P_S$ and $P_Z$ are previous loads in the load history defining the current size of the stick zone, $a(P_S,M_S)$, and the permanent stick zone, $a(P_Z,M_Z)$, respectively. These may be obtained by enforcing tangential equilibrium, see \cite{Barber_2011}. The integral in \cref{eq:CJshear} represents the accumulated shear traction in the stick zone during the transient loading, as described in detail in \cite{Barber_2011}. Although calculating the size of the stick zone, current and permanent, when the loading is purely due to $P$ and $Q$
is straightforward as demonstrated in \cite{Barber_2011,fleury2017varying}, it is not so if a moment, $M$, and a differential tension, $\sigma$, are also present.

By expanding the shear traction in a series expansion and taking only the leading-order terms, the asymptotic approximation of the shear traction at the left edge of the contact and at the minimum load point of the steady-state in Figure \ref{fig:CyclicLoop} may be obtained by a superposition of the normal bounded asymptote \cite{fleury2017varying} as
\begin{align}
q_1^{(asy)}(x) =  &K_{Q1} \sqrt{x+a_1 - d_1} + f K_{N1} \sqrt{x + a_1} - 2 f K_{N1} \sqrt{x_1 + a_1 - d_1} +\nonumber \\ 
&f K_{N1} \sqrt{x+a_2 - d_2} \text{;}
\label{eq:asymTan1}
\end{align}
%\textcolor{purple}{QUESTION: What are a and d in this equation? I assume a is the same as the end point after the transient load is completed, but what is d?}
where $d_1$ and $a_1$ are the slip zone size and contact size at the minimum load point, and $d_2$ and $a_2$ are the slip zone size and contact size at the maximum steady-state loading point. The coefficients $K_N$, here, may be obtained as in \textsection 2 and $K_Q$ is scaled from $K_N$ during the transient (and proportional) loading, i.e.~when the loading did not cause slip. 

At the maximum load point of the steady state the shear traction is given by

\begin{equation}
q_2(x) = \int_{0}^{P_Z,M_Z} \frac{\partial p(x,P,M)}{\partial P} \frac{\dd P}{\dd Q} \dd P - fp(x,P,M) + fp(x,P_Z,M_Z),
\label{eq:CJshear2}
\end{equation}%
and its analogous asymptotic approximation is
\begin{equation}
q_2^{(asy)}(x) = K_{Q2} \sqrt{x+a_2 - d_2} - f K_{N2} \sqrt{x + a_2} + f K_{N2} \sqrt{x+a_2 - d_2}.
\label{eq:asymTan2}
\end{equation}

\subsection{Calculation of the slip zone size}
\label{sec:SlipZone}

In order to calculate the size of the slip zone at the maximum and minimum points in the steady-state loading, the change in shear traction from the maximum to minimum loading points, $\Delta q$, is compared with the change of traction in the absence of slip \cite{dini2005comprehensive}. One assumption we make here is that the change in $K_N$ and $K_Q$ between the two ends of the steady-state loading can be neglected, which is valid only for moderate  loading. 
The change in shear loading  at the left hand side edge of the contact, see Figure \ref{fig:asymptotes}, from the maximum to minimum loading may thus be written as
\begin{align}
\Delta q(x) =& \;q_1^{(asy)}(x)- q_2^{(asy)}(x) \nonumber \\
=& \;K_{Q1} \sqrt{x+a_1-d_1} + f K_{N1} \sqrt{x + a_1} - 2 f K_{N1} \sqrt{x + a_1 - d_1} + f K_{N1} \sqrt{x+a_2 - d_2}  \nonumber  \\ 
&  \;- \left( K_{Q2} \sqrt{x+a_2-d_2} - f K_{N2} \sqrt{x + a_2} + f K_{N2} \sqrt{x+a_2 - d_2}\right)\; \text{.}
\end{align}

Now, let us take the limit ${(K_{N1}, K_{N2}) \rightarrow K_N}$. This simplifies the difference in shear tractions to
\begin{align}
\Delta q(x) =  f K_{N} \left( \sqrt{x+a_1} - 2\sqrt{x+a_1-d_1} + \sqrt{x+a_2 - d_2} +\sqrt{x+a_2} - \sqrt{x+a_2-d_2} \right)\;\text{.}
\end{align}

We further expand the expression at ${(x+a_1) \rightarrow \infty}$. By doing so the effects of the slip zones on the shear tractions are distant enough to justify the use of a singular traction multiplier, which yields 
\begin{align}
\Delta q(x)\simeq & f K_N \left(  2 \left( \frac{d_2}{2\sqrt{x+a_1}} + . . . \right) + \frac{\Delta a}{2\sqrt{x+a_1}} \right) \label{eq:DeltaQApprox} \\
= & \frac{\Delta K_T}{\sqrt{x + a_1}} \label{eq:DeltaQKT} \; \text{,}
\end{align}
where $\Delta a = |a_2-a_1|$ is the change in contact size over one half loading cycle at the end of the contact under consideration and where $0\leq \Delta a << a_2 + a_1$. 

We now have means of expressing the size of the slip zone at either end of the load cycle, so that by combining~\cref{eq:DeltaQKT,eq:DeltaQApprox}, the steady-state slip zone size is given in asymptotic form as 
\begin{equation}
d_1 =  -\frac{\Delta K_T}{f K_N} - \frac{\Delta a}{2}  \qquad\text{and}\qquad d_2 =  \frac{\Delta K_T}{f K_N} + \frac{\Delta a}{2}\; \text{.}
\label{eq:slipAsymptotes}
\end{equation}

We can go one step further and approximate the instantaneous size of the slip zones at every point of the steady-state load cycle. Now, the value of the asymptotic shear traction multiplier, $\Delta K_T$, must be evaluated based on the the finite changes in tangential loading, $\Delta Q = Q_2 - Q_i$ and $\Delta \sigma = \sigma_2 - \sigma_i $, where $i$ represents the instantaneous value of the respective load component. The normal traction multiplier, $K_N$, depends on the instantaneous value of the normal loading, see \textsection \ref{mossakovskii_methods}. The change in contact size is then defined as $\Delta a = |a_2- a_i|$, where $a_i$ is the instantaneous contact size.

\section{The Shallow Wedge} \label{examples}

\hspace{0.4cm}In this section we wish to draw an instructive comparison between the mathematically exact solution and its asymptotic expressions for the example of a shallow wedge subject to a cyclically varying normal load, moment, shear load and initially \textit{moderate} differential bulk tension. An example illustration of a steady-state load cycle between the minimum and maximum load points, $1$ and $2$  respectively, is given in Figure \ref{fig:CyclicLoop}. Analytical methods reach the limit of what they are able to treat when we look at problems which involve a varying moment and the differential bulk tension becomes \textit{large} enough to reverse the direction of slip at one end of the contact, see Figure \ref{fig:ModerateLargeTension} b). The asymptotic description allows us to go beyond this limitation and, here, we present the results.

\subsection{Explicit Solution}\label{explicitsolution}

\begin{figure}[t!]
	\centering
	\includegraphics[scale=0.6, trim= 0 0 0 0, clip]{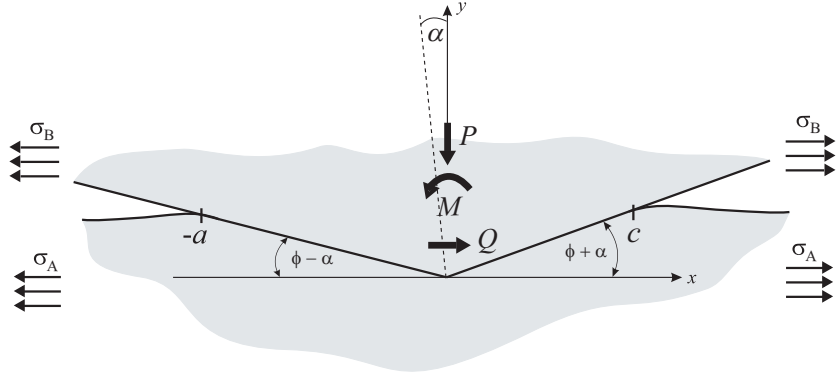}
	\caption{A tilted wedge contact subject to a normal load, moment, shear load with remote bulk tensions arising in each half-plane.}
	\label{fig:WedgeProblem}	
\end{figure}

Consider a shallow wedge of apex angle $(\pi-2\phi)$ where $0<\phi\ll1$ that is tilted at an angle $0<\alpha\ll1$ measured anti-clockwise from the unrotated state, Figure \ref{fig:WedgeProblem}. Assuming the wedge angle to be large enough for the half-plane assumption to hold, we establish the contact between the two half-planes by applying a normal force, $P$, acting through the vertex, and also subject to a moment, $M$, which is taken to be positive when $\alpha$ is positive.

The normal contact problem was first solved in \cite{Sackfield_2005}, who showed that the pressure distribution is 
\begin{equation}\label{pressure_distribution_wedge}
%p(x)= \frac{E^* \phi}{\pi}\displaystyle \ln\left\vert \frac{\sqrt{\nicefrac{1}{\gamma}} - \sqrt{\nicefrac{(a-x)}{(x+\gamma a)}}}{\sqrt{\nicefrac{1}{\gamma}}+\sqrt{\nicefrac{(a-x)}{(x+\gamma a)}}} \right\vert\;,
p(x) =  \frac{E^{*}\phi }{\pi}\log\left|\frac{\sqrt{\gamma} + \sqrt{(\gamma a-x)/(x+a)}}{\sqrt{\gamma}-\sqrt{(a\gamma-x)/(x+a)}}\right| \;,
\end{equation}
where $\gamma$ is given as \cite{Moore_2020}

\begin{equation}\label{gamma}
\gamma= \left(1-\sin\left(\frac{\pi \alpha}{2 \phi}\right)\right) \left(1+\sin\left(\frac{\pi \alpha}{2 \phi}\right)\right)^{-1}\;.
\end{equation}

The moment causes the wedge to tilt through the angle $\alpha$ which can be expressed in terms of normal load, $P$, and moment, $M$, as shown in \cite{Moore_2020} is given by

\begin{equation}\label{alpha}
\alpha = -\frac{2 \phi}{\pi} \arcsin\left(\frac{2 E^* \phi M}{\sqrt{4 (E^{*} \phi M)^2 + P^4}}\right)\;.
\end{equation}

Considering the symmetry of the wedge geometry, we expect there to be a similarity in the solution for the contact coordinates, so that
\begin{equation}
c=\gamma a \; \text{,}
\end{equation}
where $a$ can be found from normal equilibrium
\begin{equation} \label{normal_equilrbium_wedge}
P =\int_{-a}^{c}p(x)\,\mathrm{d}x =E^* \phi \sqrt{\gamma a^2}\;,
\end{equation}
to give
\begin{equation} \label{ca}
a=\frac{P}{E^* \phi}\sqrt{\frac{1}{\gamma}}\;;\qquad \qquad c=\frac{P}{E^* \phi}\sqrt{\gamma}
\;.
\end{equation}

For completeness, we also give the expression for the moment, $M$, which is
\begin{equation}
M =\int_{-a}^{c}p(x) x\,\mathrm{d}x = \frac{E^* \phi}{4} \sqrt{\gamma} (1-\gamma) a^2\;.
\end{equation}

Note that for the symmetrical \textit{no-moment} case, i.e. $M=0$, $\alpha \rightarrow 0$, $\gamma \rightarrow 1$, and $a = c$. In the case of cyclic proportional loading, as shown in Figure \ref{fig:CyclicLoop}, the normal problem will be defined in terms of the forces $P_i$ and the moments $M_i$ , $i = 1 , 2$, in which case a preliminary stage will involve the determination of the corresponding tilt angles, $\alpha_i$. Now, we state without proof the the coordinates for the permanent stick zone , $[-m \qquad n]$, as \cite{Andresen_2019_3}
\begin{equation}
m=\frac{1}{E^* \phi}\left(P_0-\frac{\Delta Q}{2f}\right)\sqrt{\frac{1}{\gamma_{\text{t}}}}\;;\qquad n=\frac{1}{E^* \phi}\left(P_0-\frac{\Delta Q}{2f}\right)\sqrt{\gamma_{\text{t}}}\;, \label{m}
\end{equation}
where $P_0$ is the mean normal load and $\Delta Q$ denotes the range of the shear load during the steady-state cycle. The function relating $m$ and $n$ is given by
\begin{equation}\label{gamma_2}
\gamma_{\text{t}}= \left(1-\sin\left(\frac{\pi \left(\alpha_0 - \frac{\Delta\sigma}{4 f E^*}\right)}{2 \phi}\right)\right) \left(1+\sin\left(\frac{\pi \left(\alpha_0 - \frac{\Delta\sigma}{4 f E^*}\right)}{2 \phi}\right)\right)^{-1}\;,
\end{equation}
where $\alpha_{0} = \nicefrac{\alpha_1 + \alpha_2}{2}$ is the average angle of tilt and $\Delta\sigma$ is the range of differential bulk tension throughout the steady-state cycle. 

With Eq. \eqref{m} we have found the extent and position of the permanent stick zone which depend on $P_0$, $\alpha_0(M_0, P_0)$, $\Delta Q$, and $\Delta\sigma$. Note that the steady-state permanent stick zone does therefore \textit{not} depend on the range of the normal load and moment, $\Delta P$ and $\Delta M$, which merely affect the contact extent at the end points of the steady-state cycle\footnote{We omit the suffix indicating the ends of the load cycle, $i = 1, 2$, for clarity.}, $[-a(P, M) ,\; c(P,M)]$. 
.

\subsection{Asymptotic Solution}

As the contact law and the permanent stick zone during the steady-state cycle are known in closed-form, as described above, the mathematically exact maximum extent of the slip zones is given as
\begin{equation}\label{slipzones_LHS}
d^{\text{LHS}}=a-m \; \text{,}
\end{equation}
at the left-hand end of the contact and
\begin{equation}\label{slipzones_RHS}
d^{\text{RHS}}=c-n \; \text{,}
\end{equation}
at the right-hand end of the contact. Here, we wish to find equivalent explicit statements using the asymptotic description outlined in \textsection \ref{asymptotic_methods}. 
In its most general form the maximum slip zone extent is given as in \cref{eq:slipAsymptotes}.

The values for $\Delta K_T$ and $K_N$ will be different at each end of the load cycle and can be found by two different means. Either we use numerical results from, say, a finite element analysis \cite{Andresen_2020,fleury16a,fleury16b}, or we apply the explicit expressions given in \textsection \ref{mossakovskii_methods} and \textsection \ref{asymptotic_methods} to write
\begin{equation}\label{K_T}
\Delta K_{T} = \frac{\Delta Q}{\pi \sqrt{2 a}} \pm \frac{\Delta\sigma}{4} \sqrt{\frac{a}{2}}\;\text{,}
\end{equation}
where we choose the upper sign when looking at the left-hand side and the lower when looking at the right-hand side. Note that the respective contact half-width ($a$, $c$) should be used for the contact edge under consideration. While $\Delta K_T$ is independent of the geometry, i.e. only the range of the tangential loads and the instantaneous contact half-width matter, the asymptotic multiplier for the normal tractions is dependent on the geometry and material of the contacting bodies. For an explicit expression, we need to know the instantaneous contact law, $P'(a)$, so that for the left-hand side of the tilted wedge we can write Eq. \eqref{eqn:Kna} in explicit form by differentiating Eq. \eqref{normal_equilrbium_wedge} with respect to $a$
\begin{equation}\label{K_N}
K_{N}^{\text{LHS}} = \frac{2 E^* \phi}{\pi}  \sqrt{\frac{\gamma}{a \, (1 + \gamma)}}\;\text{,} %\frac{2}{\pi} \frac{P'(a)}{\sqrt{a + b(a)}} = 
\end{equation}
which can be readily connected to the other end of the contact by
\begin{equation}
K_{N}^{\text{RHS}} = \frac{K_{N}^{\text{LHS}}}{\gamma} \; \text{.}
\end{equation}

\section{Display of Results}
There is a multitude of possibilities to illustrate the comparison between analytical and asymptotic results outlined above. The most sensible way seems to be to compare, first, results associated with the classical Cattaneo-Mindlin solution for periodic loading under constant normal load. We then go on to look at how this solution is affected by a varying normal load. The effects of a differential bulk tension, which may be \textit{moderate} or \textit{large}, are studied in the subsequent example. Note that for the asymptotic approach it is irrelevant whether the differential tension is moderate or large, as we only look at one end at a time. We then look at the effect of the mean angle of tilt on the solution.  Lastly, we look at the effects of a varying moment in combination with large differential tension. That is, the case in which only the asymptotic approach can provide an answer.

\subsection{The Cattaneo-Mindlin Problem}

It will be instructive to restrict ourselves, first, to the problem described in the introduction where a contact is established by a normal force, $P$, which is then held constant while a shear load, $Q$, varies between a minimum load point, $1$, and a maximum load point, $2$, over the range $\Delta Q$. This type of loading sequence was first solved for a Hertzian contact subject to a monotonically increasing tangential load \cite{Cattaneo_1938} and later extended to periodically varying tangential forces \cite{Mindlin_1951}. The J\"ager-Ciavarella theorem allows us to apply these ideas to any kind of incomplete geometry, including the shallow wedge. The established contact is symmetrical and, given the coefficient of friction, $f$, is finite, slip zones of equal extent will ensue so that the maximum slip zone extent is found using Eqs. \eqref{ca}, \eqref{m}
\begin{equation}
d = a - m =\frac{1}{E^* \phi} \left(\frac{\Delta Q}{2f}\right)\;. \label{am}
\end{equation}

 The asymptotic description yields the exact same result using Eqs. \eqref{eq:slipAsymptotes}, \eqref{K_T}, and \eqref{K_N}
\begin{equation}
d = \frac{\Delta K_T}{f K_N} =\frac{1}{E^* \phi} \left(\frac{\Delta Q}{2f}\right)\;. \label{d}
\end{equation}

\begin{figure}[t!]
	\centering
	\includegraphics[scale=0.4, trim= 0 0 0 0, clip]{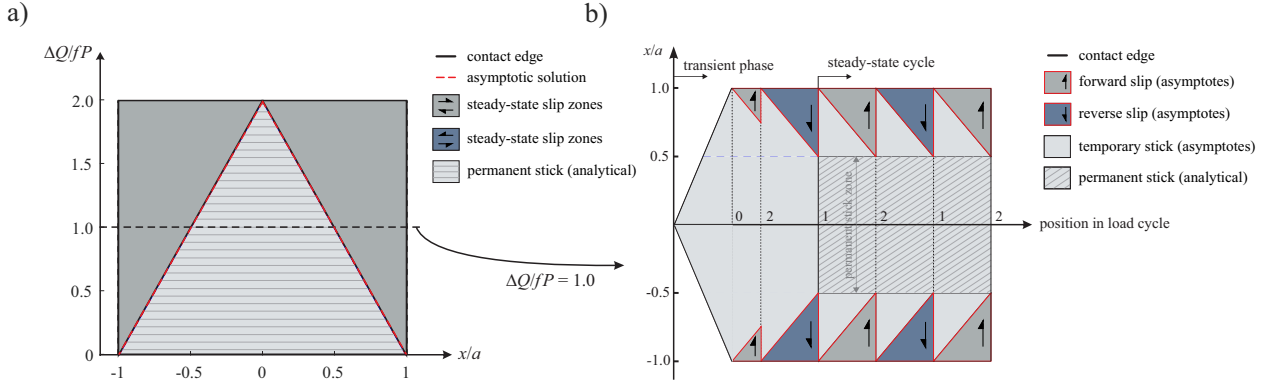}
	\caption{a) Stand-alone steady-state solutions under constant normal load showing the contact size and maximum slip zone extent for $ 0 < \nicefrac{\Delta Q}{f P} < 2$, b) the marching-in-time solution for $\nicefrac{\Delta Q}{f P} = 1.0$ with zones of forward and reverse slip.}
	\label{fig:DeltaQ_Steady_state}	
\end{figure}

The reason for this is the proportionality between applied normal load and contact size of the wedge solution, see Eq. \eqref{ca}. For a different geometry, a relative error between the exact and asymptotic solutions is expected, which will depend on the nature of the contact law. Figure \ref{fig:DeltaQ_Steady_state} a) shows the linearity of the steady-state solution associated with this problem for vales of $ 0 < \nicefrac{\Delta Q}{f P} < 2$. For each value of $\nicefrac{\Delta Q}{f P}$ a stand-alone maximum slip zone size extent can be obtained. In Figure \ref{fig:DeltaQ_Steady_state} b) the steady-state solution for $\nicefrac{\Delta Q}{f P} = 1.0$ is plotted in a marching-in-time sense. After a transient phase - note, the path taken to reach the steady-state only affects the locked-in tractions in the permanent stick zone, but not the steady-state slip-stick pattern - the steady-state cycle is established and we see zones of reciprocating slip emerge at both ends of the contact. As said before, since we have chosen the wedge geometry, the Cattaneo-Mindlin problem yields the exact same solution as the asymptotic and analytical approach for the maximum extent of the slip zones, see Eqs. (\ref{am} and \ref{d}). Note that the explicit solution, Eq. \eqref{am}, will allow us to find the maximum extents of the steady-state slip zones only, while the asymptotic solution makes it possible to \textit{march} through the steady-state load cycle and find the slip-stick boundary at every point in time. For the sake of comparability, we will adopt this style of presenting the results for the other loading scenarios considered in this work.

\subsection{The Effect of a Varying Normal Load}

\begin{figure}[t!]
	\centering
	\includegraphics[scale=0.4, trim= 0 0 0 0, clip]{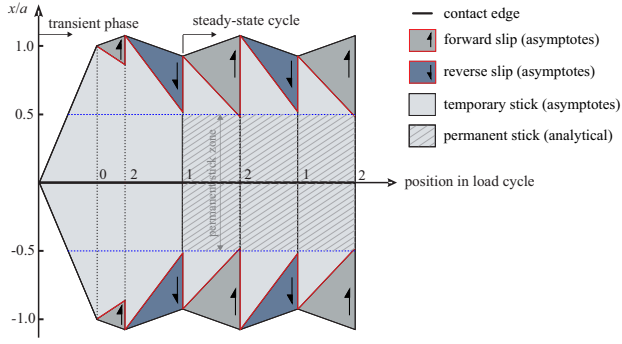}
	\caption{Marching-in-time solution for a varying normal load problem,  $\nicefrac{\Delta P}{P} = 0.15$, while the tangential load variation is $\nicefrac{\Delta Q}{f P} = 1.0$.}
	\label{fig:DeltaP_Steady_state}	
\end{figure}

From Eqs. (\ref{m} and \ref{gamma_2}) we see that the position and extent of the permanent stick zone is independent of variations in normal load, $\Delta P$. The slip zone size will vary, of course, due to the variation in size of the contact. The asymptotic description of the slip zone size, Eq. \eqref{eq:slipAsymptotes}, depends on the instantaneous values of the asymptotic multipliers, $\Delta K_T$ and $K_N$, as well as the overall change in contact size, $\Delta a$, throughout a cycle. This means there is no explicit dependence on the normal load variation, $\Delta P$. Implicitly, however, all three of the stated parameters are affected by changes in normal load. It is therefore of interest to employ the asymptotic framework developed to a varying normal load problem and demonstrate its applicability. In order to simplify comparability, we will use the same parameters as used in the example above, i.e. $\nicefrac{\Delta Q}{f P} = 1.0$ and add a varying normal load of $\nicefrac{\Delta P}{P} = 0.15$ to the problem. Figure \ref{fig:DeltaP_Steady_state} shows a marching-in-time solution for the problem described. We notice that the size and position of the permanent stick zone remains unchanged compared with Figure \ref{fig:DeltaQ_Steady_state} b). The overall contact size, however, increases and decreases throughout the cycle, therefore effectively changing the size of the slip zones not only due to a variation in shear load, but also due to movement of the contact edge. The asymptotic framework tends to underestimate the slip zone size at the minimum load point of the cycle and to overestimate it at the maximum load point of the cycle. While it is not possible to make a general statement regarding the relative error, we regard the results obtained for this example case as acceptable, since the error is certainly less than $5\%$. Encouraged by this comparison, in the examples which follow, we will refrain from adding the effect of a varying normal load, in order to concentrate on the effects of the loading under consideration.

\subsection{The Effect of Differential Bulk Tension}
Bulk tensions are stresses arising remotely from the contact interface in one or both contacting bodies. In an engineering application like a gas turbine engine, centrifugal loading acting on the components or vibrations might be a source of remote tensions. These cause differential tensions between the contacting components exciting interfacial shear tractions. Note, that in most practical applications the tangential loading is a mix of differential tension and a shear force. Here, we look at what happens to the steady-state solution when in addition to a varying shear force, $\Delta Q$, a varying differential tension, $\Delta\sigma$ is present. We will use the results presented in \textsection \ref{explicitsolution} to compare the closed-form solution for moderate differential tension to the asymptotic description. The solution given in Eq. \eqref{m} is valid as long as the same sign slip direction is maintained at both ends of the contact. That is, the range of differential tension, $\Delta\sigma$, is smaller than or equal to its transition value to the onset of reverse slip, $\Delta\sigma_{th}$,
\begin{align}\label{inequalitywedge}
\Delta\sigma \leq \Delta\sigma_{th}=\frac{4 E^* f}{\pi} \left[\pi \alpha_0  - 2  \phi \sin^{-1} \left(\frac{8 f^2 P^2}{ \Delta Q \gamma (- 4 f P+\Delta Q)+(1+\gamma) 4 f^2 P^2}-1\right)\right]\;.
\end{align}

Suppose now, that the range of differential tension is \textit{large} enough for inequality \eqref{inequalitywedge} to be violated, $\Delta\sigma > \Delta\sigma_{th}$, so that the slip directions are of opposite sign as shown in Figure \ref{fig:ModerateLargeTension} b). In order to have a mathematically exact point of comparison against the asymptotic solution, we use the following consistency conditions for constant normal load problems subject to varying \textit{large} differential tension and shear load from Andresen et al. \cite{Andresen_2019_4} and evaluate $m$ and $n$ numerically
\begin{align}
\frac{\Delta\sigma\pi}{8}= & \,-\int_{-a}^{-m}\frac{fp(\xi)\,\text{sgn}(\xi)}{\sqrt{(\xi-n)(\xi+m)}}\,\mbox{d}\xi+\int_{n}^{a}\frac{fp(\xi)\,\text{sgn}(\xi)}{\sqrt{(\xi-n)(\xi+m)}}\,\mbox{d}\xi,\label{eq17}\\
\frac{\Delta Q}{2}+\frac{(n-m)\Delta\sigma\pi}{16}= & \,-\int_{-a}^{-m}\frac{\xi fp(\xi)\,\text{sgn}(\xi)}{\sqrt{(\xi-n)(\xi+m)}}\,\mbox{d}\xi+\int_{n}^{a}\frac{\xi fp(\xi)\,\text{sgn}(\xi)}{\sqrt{(\xi-n)(\xi+m)}}\,\mbox{d}\xi.\label{eq18}
\end{align} 

\begin{figure}[t!]
	\centering
	\includegraphics[scale=0.4, trim= 0 0 0 0, clip]{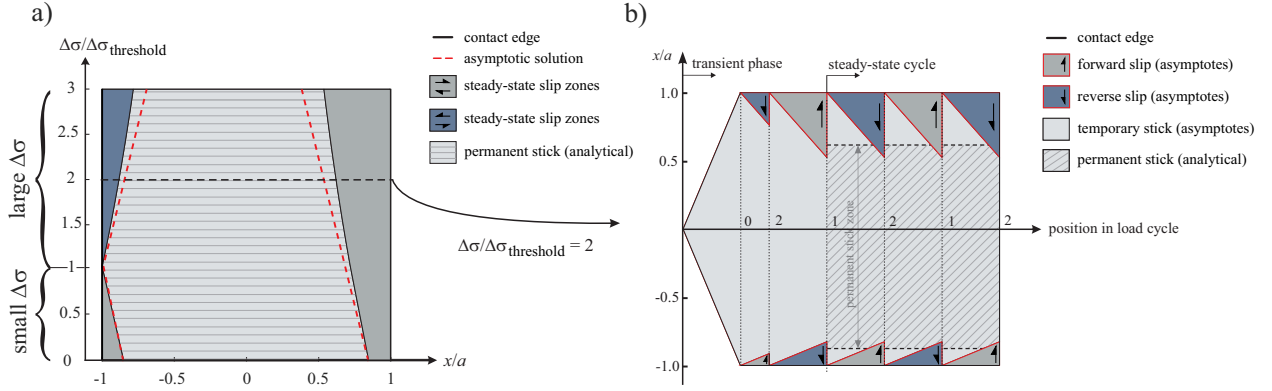}
	\caption{a) Stand-alone steady-state solutions under constant normal load showing the contact size and maximum slip zone extent for $\Delta\sigma/\Delta\sigma_{th} < 3$, b) the marching-in-time solution for $\Delta\sigma/\Delta\sigma_{th} = 2$ with opposing zones of slip at both ends of the contact.}
	\label{fig:DeltaSigma}	
\end{figure}

A closed-form solution can be obtained for the Hertzian geometry, and we provide this in appendix \ref{explicitsolutionHertz} for reference. In the example we examine in this section, the contact is established by the normal load, $P$ until the contact half-width reaches $a$. The shear force is then varied with range $\Delta Q = 0.3 f P$ and we see the effect of a change in differential bulk tension over $0 \leq \Delta\sigma/\Delta\sigma_{th} < 3$ for the asymptotic and analytical solutions, Figure \ref{fig:DeltaSigma} a). Note that the $\Delta\sigma$ is normalised to its transition value, $\Delta\sigma_{th}$, see Eq. \eqref{inequalitywedge}. When the transition value is reached, $\Delta\sigma/\Delta\sigma_{th} = 1$, we observe that the direction of slip changes at one end of the contact and the underlying analytical approach switches from the explicit branch of its solution for moderate differential tension, Eq. \eqref{m}, to the numerical branch, Eqs. (\ref{eq17} and \ref{eq18}) for large differential tension. The larger the slip zones become relative to the overall contact size, the less accurate the asymptotic description becomes. We will refrain from further assessing the relative error between analytical and asymptotic approach as the number of input parameters is too large to make any general statements beyond the one we have just made. It is the responsibility of the user to acknowledge this trade-off between the mathematically exact analytical approach and the very convenient asymptotic description. 

Figure \ref{fig:DeltaSigma} b) on the other hand shows the steady-state slip-stick pattern in a marching-in-time sense for large differential tension, $\Delta\sigma/\Delta\sigma_{th} = 2$. Note that the direction of slip is of opposite sign at each end of the contact and that the asymptotic description overestimates the maximum extent of the slip zones at both ends of the contact.

\subsection{The Effect of Tilting}

\begin{figure}[t!]
	\centering
	\includegraphics[scale=0.4, trim= 0 0 0 0, clip]{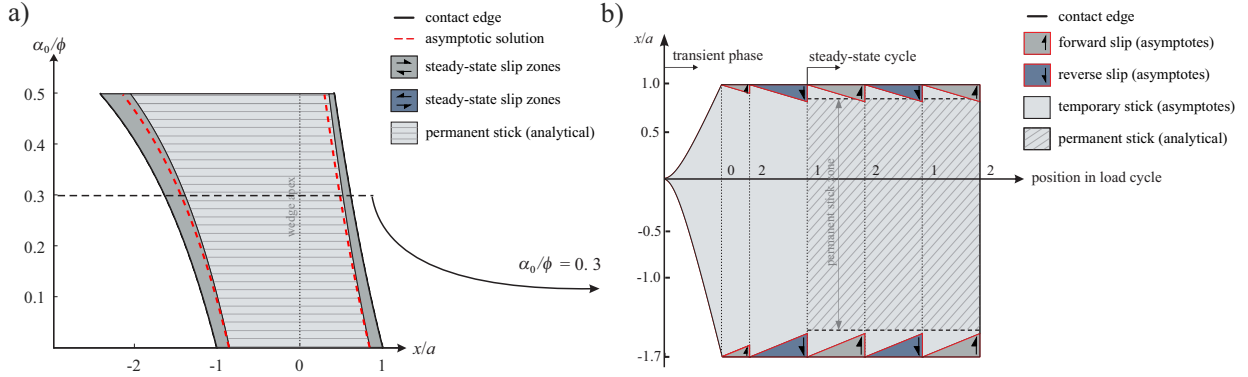}
	\caption{a) Stand-alone steady-state solutions under constant normal load showing the contact size and maximum slip zone extent for $ 0 < \alpha_0/\phi < 0.5$ subject to a varying shear force, $\Delta Q/f P0 = 0.3$, b) the marching-in-time solution for $\alpha_0/\phi = 0.3$ with zones of forward and reverse slip.}
	\label{fig:evolution_of_slipstick_with_alpha}	
\end{figure}

In most engineering applications, there is a moment transmitted through the interface causing the bodies to tilt relative to each other. From \textsection \ref{explicitsolution} we know that the coordinates for the permanent stick zone $[-m(\alpha_0) \qquad n(\alpha_0)]$ change with the average angle, $\alpha_0$, unaffected by the actual range of the applied moment, $\Delta M$. Consider Figure \ref{fig:evolution_of_slipstick_with_alpha} a). Note that each value of $\alpha_0$ represents a stand-alone steady-state solution subject to a constant normal load, $P_0$, giving the mean contact size spanning $[-a_0, \qquad c_0]$. In this example, we keep the shear load fluctuation, $\Delta Q = 0.3 f P_0$ constant, giving the permanent stick zone spanning $[-m, \quad n]$.  
The fluctuation in bulk stress, $\Delta\sigma$, is kept zero as the intention is to demonstrate the effects of tilt exclusively. Note that once a differential tension is applied and it becomes large enough to reverse the direction of slip at one end of the contact, the analytical solution used here as a point of comparison becomes inapplicable and only the asymptotic approach provides an answer. From Figure \ref{fig:evolution_of_slipstick_with_alpha}, it is apparent that increasing the average angle of tilt influences not only the extent and position of the contact, but also the extent and position of the permanent stick zone within the contact. The example reflects the solution behaviour indicated by Eq. \eqref{m}. We note that it is the mean normal load, $P_0$, and the average angle of tilt, $\alpha_0$, which affect the tangential solution, together with the tangential load inputs, $\Delta\sigma$ and $\Delta Q$. Here, we have not looked at the effects of the latter two, $\Delta\sigma$ and $\Delta Q$, but the solution developed above will remain valid as long as a shear-dominant behaviour is expected, i.e. the change in differential bulk tension is such that the direction of slip is the same at both ends of the contact during each half-cycle, and the permanent stick zone is in place at both sides of the wedge apex, i.e. the coordinates, $m$ and $n$, remain positive.

Figure \ref{fig:evolution_of_slipstick_with_alpha} b) shows the marching-in-time steady-state solution with the expected asymmetrical slip-stick behaviour across the interface. Note that if the average angle of tilt, $\alpha_0$, becomes zero, a symmetrical slip-stick pattern, qualitatively equal to that presented in Figure \ref{fig:DeltaQ_Steady_state}, would be recovered. In the example illustrated in Figure \ref{fig:evolution_of_slipstick_with_alpha}	b), the asymptotic solution underestimates the size of the steady-state slip zone at one end, while overestimating it at the other end. Overall, however, we see good agreement between the two different approaches for all depicted values of $\alpha_0/\phi$.

\subsection{A varying moment together with large differential tension}

\begin{figure}[t!]
	\centering
	\includegraphics[scale=0.4, trim= 0 0 0 0, clip]{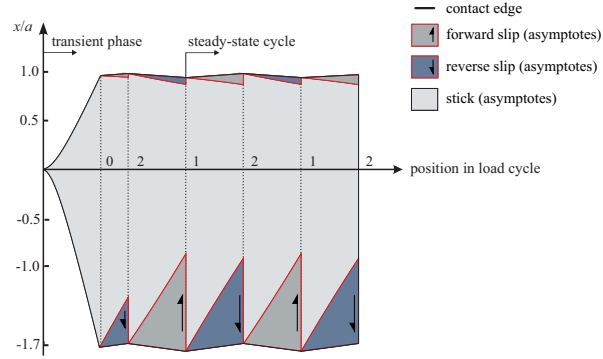}
	\caption{The marching-in-time solution for a shallow wedge with a normalised angle of tilt, $\alpha/\phi$, varying between $0.28$ and $0.32$ while being subject to mixed shear loading of $\Delta Q/f P0 = 0.3$ and large differential tension, $\Delta\sigma/\Delta\sigma_{th} = 2$ with zones of forward and reverse slip using the asymptotic description.}
	\label{fig:varyingmoment}	
\end{figure}

In all loading scenarios examined so far in this paper there has always been a mathematically exact solution for the maximum extent of the steady-state slip zones with which to compare their asymptotic description. As explained in the introduction, the analytical approach reaches its limitation once we consider a \textit{large} varying bulk tension in the presence of a varying moment, or angle of tilt. Now that we have gained confidence in the applicability of the asymptotic description, we fill-in this remaining puzzle piece by revisiting the case of Figure \ref{fig:evolution_of_slipstick_with_alpha} b). Now the normalised angle of tilt, $\alpha/\phi$, is going to vary between $0.28$ and $0.32$, as opposed to being kept constant at $0.3$. This is going to cause a repetitive movement of the contact edge throughout the cycle. In addition to varying shear force, $\Delta Q/f P0 = 0.3$, we apply a large differential tension, $\Delta\sigma/\Delta\sigma_{th} = 2$, so that the direction of slip is not the same at both ends of the contact. 
Figure \ref{fig:varyingmoment} shows the marching-in-time slip-stick pattern throughout the cycle. Note that while the slip zones at one end are small compared with the overall contact half-width, they advance quite significantly into the inner part of the contact at the other end of the contact. From the experience gained with previous examples we know that the smaller the slip zones compared with the overall contact size, the higher the quality of approximation, so we may expect the asymptotic approach to eventually break down at one end of the contact.

\section{Conclusions}

In this contribution we have outlined recipes for obtaining the slip-stick behaviour for incomplete contacts under oscillatory loading of normal load, moment, shear load and differential bulk tension using asymptotic and analytical approaches. The motivation for this work stems from the limitations of the analytical description, that is, the lack of a mathematically exact solution to the steady-state slip stick behaviour in the presence of a varying moment and a differential bulk tension large enough to reverse the direction of slip at one end of the contact. A comparison is drawn between the asymptotes and the analytical solution for a shallow wedge geometry for different loading scenarios in order to gain confidence in the asymptotic approach. The work concludes with the application of the asymptotic description to a case where no existing analytical approach can provide an answer. The asymptotic approach presented herein equips the user with a simple-to-apply methodology for tackling problems for which these analytical problems manifest. The methods will also be of significant practical use in real-world problems where the complexity of the geometry may preclude analytical progress

\section*{Acknowledgements}
\hspace{0.4cm}This project has received funding from the European Union's Horizon 2020 research and innovation programme under the Marie Sklodowska-Curie agreement No 721865. David Hills thanks Rolls-Royce plc and the EPSRC for the support under the Prosperity Partnership Grant “Cornerstone: Mechanical Engineering Science to Enable Aero Propulsion Futures”, Grant Ref: EP/R004951/1.

\bibliography{References_Jan2020}

\begin{thebibliography}{10}
\expandafter\ifx\csname url\endcsname\relax
  \def\url#1{\texttt{#1}}\fi
\expandafter\ifx\csname urlprefix\endcsname\relax\def\urlprefix{URL }\fi
\expandafter\ifx\csname href\endcsname\relax
  \def\href#1#2{#2} \def\path#1{#1}\fi

\bibitem{Cattaneo_1938}
C.~Cattaneo, {Sul Contato di Due Corpo Elastici}, Atti Accad. Naz. Lincei, Cl.
  Sci. Fis., Mat. Nat., Rend. 27 (1938) 342--348, 434--436, 474--478.

\bibitem{Jaeger_1997}
J.~J\"ager, {Ein neues Prinzip der Kontaktmechnik}, ZAMM 77, Issue S1 (1997)
  143--144.

\bibitem{Ciavarella_1998}
M.~Ciavarella, The generalised cattaneo partial slip plane contact problem,
  part i theory, part ii examples, Int. Jnl. Solids Struct. 35 (1998)
  2349--2378.

\bibitem{Hills_1987}
D.~Nowell, D.~Hills, {Mechanics of fretting fatigue tests}, Int. Jnl. Mech.
  Sci. 29 (1987) 355--365.

\bibitem{Barber_2011}
J.~Barber, M.~Davies, D.~Hills, Frictional elastic contact with periodic
  loading, Int. Jnl. Solids Struct. 48 (2011) 2041--2047.

\bibitem{Andresen_2019_3}
H.~Andresen, D.~Hills, J.~Barber, J.~V\'azquez, Steady state cyclic behaviour
  of a half-plane contact inpartial slip subject to varying normal load,
  moment, shear load, and moderate differential bulk tension, Int. Jnl. Solids
  Struct. accepted for publication.

\bibitem{Andresen_2019_2}
H.~Andresen, D.~Hills, J.~Barber, J.~V\'azquez, Frictional half-plane contact
  problems subject to alternating normal and shear loads and tension in the
  steady state, Int. Jnl. Solids Struct. 168 (2019) 166--171.

\bibitem{Moore_2018}
M.~R. Moore, R.~Ramesh, D.~Hills, J.~Barber, Half-plane partial slip contact
  problems with a constant normal load subject to a shear force and a
  differential bulk tension, Jnl. Mech. Phys. Solids 118 (2018) 245--253.

\bibitem{Andresen_2019_4}
H.~Andresen, D.~Hills, M.~Moore, Steady state partial slip problem for half
  plane contacts subject to a constant normal load using glide dislocations,
  Eur. Jnl. Mech/A Solids 79.

\bibitem{Andresen_2020}
H.~Andresen, M.~Moore, D.~Hills, Representation of incomplete contact problems
  by half-planes, Eur. Jnl. Mech/A Solids (2020) under review.

\bibitem{Moore_2020}
M.~R. Moore, D.~A. Hills, {Extending the Mossakovskii method to contacts
  supporting a moment}, J. Mech. Phys. Solids. 141 (2020) 103989.

\bibitem{Barber_2010}
J.~Barber, Elasticity (third edition), Springer, 2010.

\bibitem{Mossakovskii_1953}
V.~I. Mossakovskii, Application of the reciprocity theorem to the determination
  of the resultant forces and moments in three-dimensional contact problems,
  PMM 17 (1953) 477--482.

\bibitem{Dundurs_1969}
J.~Dundurs, {Discussion of edge-bonded dissimlar orhtogonal wedges under normal
  and shear loading (by D. Bogey)}, Jnl. Appl. Mech. 36 (1969) 342--348, 650 --
  652.

\bibitem{sackfield2003application}
A.~Sackfield, A.~Mugadu, J.~R. Barber, D.~A. Hills, The application of
  asymptotic solutions to characterising the process zone in almost complete
  frictionless contacts, J. Mech. Phys. Solids 51~(7) (2003) 1333--1346.

\bibitem{dini2004bounded}
D.~Dini, D.~A. Hills, Bounded asymptotic solutions for incomplete contacts in
  partial slip, Int. J. Solids Struct. 41~(24) (2004) 7049--7062.

\bibitem{dini2005comprehensive}
D.~Dini, A.~Sackfield, D.~A. Hills, Comprehensive bounded asymptotic solutions
  for incomplete contacts in partial slip, J. Mech. Phys. Solids 53~(2) (2005)
  437--454.

\bibitem{Jaeger_1998}
J.~J\"ager, A new principle in contact mechanics, Jnl. Tribology 120 (1998)
  677--684.

\bibitem{fleury16a}
R.~M.~N. Fleury, D.~A. Hills, J.~R. Barber, A corrective solution for finding
  the effects of edge rounding on complete contact between elastically similar
  bodies. part i: Contact law and normal contact considerations, Int. J. Solids
  Struct. 85 - 86 (2016) 89 -- 96.

\bibitem{fleury16b}
R.~M.~N. Fleury, D.~A. Hills, J.~R. Barber, A corrective solution for finding
  the effects of edge rounding on complete contact between elastically similar
  bodies. part ii: Near-edge asymptotes and the effect of shear, Int. J. Solids
  Struct. 85-86 (2016) 97 -- 104.

\bibitem{fleury2017varying}
R.~M.~N. Fleury, D.~A. Hills, R.~Ramesh, J.~R. Barber, Incomplete contacts in
  partial slip subject to varying normal and shear loading, and their
  representation by asymptotes, J. Mech. Phys. Solids 99 (2017) 178 -- 191.

\bibitem{hills16}
D.~A. Hills, R.~M.~N. Fleury, D.~Dini, Partial slip incomplete contacts under
  constant normal load and subject to periodic loading, Int. J. Mech. Sci.
  108-109 (2016) 115 -- 121.

\bibitem{Sackfield_2005}
A.~Sackfield, D.~Dini, D.~Hills, The tilted shallow wedge problem, European
  Journal of Mechanics A/Solids 24 (2005) 919--928.

\bibitem{Mindlin_1951}
R.~Mindlin, W.~Mason, T.~Osmer, H.~Dereiewicz, Effect of an oscillating
  tangential force on the contact surfaces of elastic spheres, Proc. First Nat.
  Congress of Applied Mechanics (1951) 203--208.

\bibitem{Hertz_1881}
H.~Hertz, {\"Uber die Ber\"uhrung fester elastischer K\"orper}, Journal f\"ur
  die reine und angewandte Mathematik 92 (1881) 156--171.

\end{thebibliography}

\newpage
\appendix
\section{Appendix} \label{Appendix}

%\subsection{Slip zone size using the asymptotic approach}
%
%In order to calculate the size of the slip zone at point B, we need to write the change in shear traction during this step (i.e.~Eq.~(26) in JMPS paper), however here I neglect the contribution of $K_Q$, I tried once to include it, but the series expansion to obtain $d_B$ became too complicated.
%\begin{equation}
%\Delta q_{BC} = f K_N^C \left( \sqrt{x+a_B} - 2\sqrt{x+a_B-d_B} + \sqrt{x+a_C - d_C} \right) - f K_N^C \left( -\sqrt{x+a_C} + \sqrt{x+a_C-d_C} \right).
%\end{equation}
%%
%Using the approximation $K_N^C \simeq K_N^B$, valid for small changes of contact size,
%\begin{eqnarray}
%\frac{\Delta q_{BC}}{f K_N^B} &=&  \left( \sqrt{x+a_B} - 2\sqrt{x+a_B-d_B} + \sqrt{x+a_C - d_C} \right) - \left( -\sqrt{x+a_C} + \sqrt{x+a_C-d_C} \right) \\
%& = & \sqrt{x+ a_B} - 2 \sqrt{x+a_B-d_B} + \sqrt{x+a_B + \Delta a}, \\
%& \simeq &  \sqrt{x+ a_B} - 2 \sqrt{x+a_B-d_B} + \left ( \sqrt{x+a_B} + \frac{\Delta a}{2 \sqrt{x+a_B}} + . . .  \right), \\
%& \simeq & 2 \left( \sqrt{x+ a_B} - \sqrt{x+a_B-d_B}\right) + \frac{\Delta a}{2\sqrt{x+a_B}}, \\
%& \simeq & 2 \left( \frac{d_B}{2\sqrt{x+a_B}} + . . . \right) + \frac{\Delta a}{2\sqrt{x+a_B}}.
%\end{eqnarray}
%Since we also have that $\Delta q_{BC} \simeq K_T/\sqrt{x+a_B}$, we can combine it with equation 19 to obtain the size of the slip zone
%\begin{equation}
%d_B = \frac{\Delta K_T}{f K_N^B} + \frac{\Delta a}{2}
%\end{equation}%

\subsection{Analytical Solution - Hertzian contact}\label{explicitsolutionHertz}
The simplest problem which may be treated is that of two cylinders of relative curvature $1/R$
pressed together by a normal force $P$ over $[-a \qquad a]$ with their axes parallel. This is the two-dimensional half-plane
equivalent of the solution first found by Hertz \cite{Hertz_1881}. The pressure distribution is given as
\begin{equation}\label{pressure_distribution_Hertz}
p(x)= p_0 \sqrt{a^2-x^2}\;,
\end{equation}
where $p_0 = \nicefrac{E^*}{2 R}$. The contact law is found by evaluating normal equilibrium
given by 
\begin{align}\label{Hertz_contactlaw}
P = \int_{-a}^{a} p(x) \,\mathrm{d}x=\frac{a^2 \pi E^*}{4 R} \, \text{,}
\end{align}
so that the half-width of the
contact patch varies with $\sqrt{P}$
\begin{align}
a^{2}=\frac{4PR}{\pi E^*}\; \text{.}
\end{align}

Let us consider the tangential problem of an applied varying shear force over range $\Delta Q>0$, see Figure \ref{fig:CyclicLoop}. The analytical solution to the partial slip problem is found by evaluation of tangential equilibrium so that the steady-state permanent stick zone has extent \cite{Andresen_2019_2}
\begin{align}\label{Hertz_permanent_stick_zone} 
d^{2}=\frac{4R}{\pi E^*}\left[  P_{0}-\frac{\Delta Q}{2 f}\right] \, \text{,}
\end{align}
where $P_0$ is the mean normal load associated with the mid-point of the steady-state load cycle. In the presence of a simultaneously exerted varying \textit{moderate} differential tension of range $\Delta \sigma>0$, the permanent stick zone shifts from its central positioning with eccentricity 
\begin{align}\label{Hertz_shift_stick_zone}
e=-\frac{R\Delta\sigma}{4fE^*} \, \text{.}
\end{align}

Now the permanent stick zone spans $[-m \qquad n]$, where $-m=-d+e$ and $n=d+e$. It is noteworthy that here, in the simple case of a Hertzian the geometry subject to \textit{moderate} mixed tangential loading, the extent and position of the permanent stick zone are given as a pair of uncoupled explicit equations. If the geometry profile gradient is different from a Hertzian contact, i.e. is not a linear function, the two equations might be coupled and the size and the eccentricity of the permanent stick region are to be obtained implicitly.

The solution given in Eqs. (\ref{Hertz_permanent_stick_zone}, \ref{Hertz_shift_stick_zone}) is valid as long as the same sign slip direction is maintained at both ends of the contact and the change in differential tension satisfies \cite{Andresen_2019_4}
\begin{align}\label{inequality}
\Delta\sigma \leq \frac{4 E^*f}{R} \left(a-\sqrt{a^2-\frac{2 R \text{$\Delta $Q} }{\pi E^* f}}\right)\;.
\end{align}

Suppose now, that the range of differential tension is \textit{large} enough for inequality \eqref{inequality} to be violated so that the slip directions are of opposite sign as shown in Figure \ref{fig:ModerateLargeTension} b). From Andresen et al. \cite{Andresen_2019_4} we know that $m$ and $n$ must satisfy the following non-linear implicit equations
\begin{align}\label{eq25}
\frac{\Delta\sigma\pi}{8} =&\frac{E^*f}{2 R}\left[\alpha_{1}\mbox{E}(\chi)+\beta_{1}\mbox{K}(\chi) + \gamma_{1}\Pi\left(\phi,\chi\right) + \delta_{1}\Pi\left(\omega,\chi\right)\right],
\end{align}
\begin{align}\label{eq26}	
\frac{\Delta Q}{2} + \frac{(n-m)\Delta\sigma\pi}{16}=&\frac{E^*f}{2 R}\left[\alpha_{2}\mbox{E}(\chi)+\beta_{2}\mbox{K}(\chi) + \gamma_{2}\Pi\left(\phi,\chi\right) + \delta_{2}\Pi\left(\omega,\chi\right)\right],
\end{align}
where $\mbox{K}(\chi)$, $\mbox{E}(\chi)$, and $\Pi(t,\chi)$ are elliptic integrals of the first, second and third kind and
\begin{alignat*}{5}
\phi &= && \frac{m-a}{m+a}, &&\; \omega &&= &&\frac{a-n}{a+m}, \\
\alpha_{1} & = && -2\sqrt{(a+m)(a+n)}, && \; \alpha_{2} && = && \frac{3(m-n)}{2}\sqrt{(a+m)(a+n)}, \\ 
\beta_{1} & = && (n-m+4a)\sqrt{\frac{a+m}{a+n}}, && \; \beta_{2} && = && \frac{-3}{2}\sqrt{\frac{a+m}{a+n}}\left(\left(\frac{n}{3}+2a\right)m-\frac{1}{2}(m^{2}+n^{2})-\frac{2a}{3}(a+n)\right), \\
\gamma_{1} & = && \frac{2a(n-m)}{\sqrt{(a+m)(a+n)}}, && \; \gamma_{2} && = && \frac{3a}{2\sqrt{(a+m)(a+n)}}\left(n^{2}-\frac{2mn}{3}+m^{2}-\frac{4a^{2}}{3}\right), \\
\delta_{1} & = && \frac{m^{2}-n^{2}}{\sqrt{(a+m)(a+n)}}, && \; \delta_{2} && = && \frac{-3(m+n)}{4\sqrt{(a+m)(a+n)}}\left(n^{2}-\frac{2mn}{3}+m^{2}-\frac{4a^{2}}{3}\right), \,\text{and}\\
\chi & = && \frac{\sqrt{(a^{2}-m^{2})(a^{2}-n^{2})}}{(a+m)(a+n)}. && && &&
\end{alignat*}

Equations \eqref{eq25} and \eqref{eq26} are sufficient to define the size and position of the permanent stick zone in the case of \textit{large} differential tension. From Eq. \eqref{eqn:Kna} we know that the asymptotic multiplier is dependent on the geometry-specific contact law and for a Hertz-type contact we write
\begin{equation}
K_{N} = \frac{E^*}{2 R}\,\sqrt{2 a} \; \text{.}
\end{equation}

These results are sufficient to carry out any type of asymptotic analysis outlined in this work and to compare against the analytical description where applicable.

\end{document}